# Versatile, open-access opto-mechanics platform for optical microscopes prototyping

Łukasz Zinkiewicz, Milena Królikowska, Alexander Krupiński-Ptaszek, Piotr Wasylczyk*

Photonic Nanostructure Facility, Faculty of Physics, University of Warsaw, ul. Pasteura 5, Warsaw 02-093, Poland

*corresponding author: pwasylcz@fuw.edu.pl

keywords: optical microscope, opto-mechanics, microscope mechanics, microscope platform

**ABSTRACT**

Prototype optical microscopes, built to pursue developments in advanced imaging techniques, need specific optomechanical constructions: preferably with high flexibility in the elements' arrangement, easy access to the optical paths, straightforward integration with external optical subsystems – light sources and detectors – as well as good mechanical stability. Typically they are either built around an adapted commercial microscope body or as a home-built setup, based on standard optomechanical elements, and neither solution delivers the desired characteristics. We developed a series of versatile platforms for prototyping optical microscopes in various configurations that use folding mirror(s) to maintain the optical paths horizontal throughout most of the setup, thus enabling the use of standard optical components in the excitation and detection paths and, last but not least, increasing the laser safety of the optical system.

**INTRODUCTION**

After a period of relative stagnation, many new ideas emerged in optical microscopy around the turn of the century. As a result, it is now possible to achieve astonishing resolutions, well beyond the Abbe limit [1] and image whole organs and entire organisms with cellular resolution [2]. This spectacular progress has been enabled by exploring previously ignored light-matter interactions, such as nonlinear light-matter interactions in STED microscopy [3], using clever optical configurations, like in light sheet microscopy [4], newly developed components, such as super-sensitive cameras in SOFI [5], data processing, as in STORM [6], or combinations of these. As a result, a contemporary super-resolving optical microscope has little in common with its 50 years old predecessor in terms of the optical setup, sample illumination, signal acquisition and processing. Yet, the opto-mechanics, the microscope "body" in particular, looks strikingly similar to the early designs, with perhaps the biggest difference in the case of the inverted configuration, developed to offer better access from the top to the sample plane.

Early optical microscopes were built with the sample laying on a horizontal stage and vertical light path, going through the objective and the eyepiece, the latter arranged conveniently for a user sitting or standing at the desk – Fig. 1. Today, eyepieces have been replaced with high resolution and high dynamic range cameras, also due to safety concerns with laser illumination being used in many microscopes. Illumination is no longer provided with concentrated sunlight or light bulbs, but rather by sophisticated LED or laser systems, the latter, e.g. in the case of femtosecond lasers for two- or three-photon excitation, often being larger in size than the microscope itself.

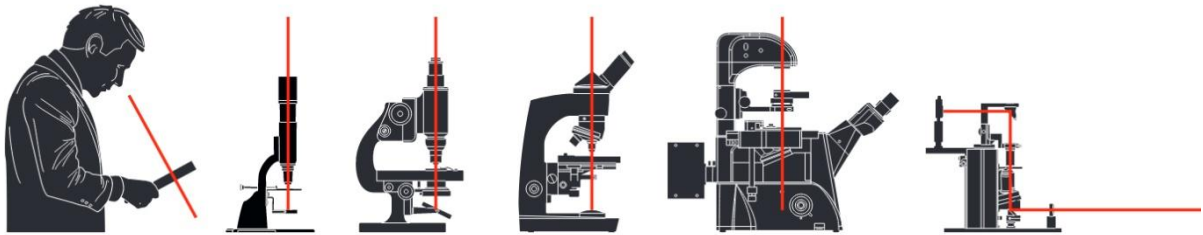

Figure 1 | From handheld magnifying glass to the contemporary inverted multi-modal imaging workstation – the evolution of the optical microscope optical layout. For good reasons the preferred orientation of the sample – especially in the case of a wet environment – is horizontal and thus the natural orientation of the principal optical axis is vertical. The last panel presents the schematic of the idea of the horizontal microscope platform, where light travels vertically over a very short distance only, between the illumination and collection optics (objectives).

Apart from the majority of microscope users, who simply want the device best suited for their application, there is a much smaller number of researchers who build or modify their microscopes, usually to go beyond what is possible with commercially available equipment [7]. The modifications in and around the microscope body can include using non-standard light sources and/or detectors, inserting additional optical elements in the excitation and/or detection light paths, or reconfiguring the sample holder. One quickly realizes that when using a commercial microscope body, it is often very inconvenient to access certain points along the optical path – sometimes additional relay optics may be added to address this problem. It is not unlikely that what remains in the new construction from the sophisticated (and expensive) inverted microscope is the nosepiece and the focusing block and all the other components, mechanical and optical, are added, often in awkward locations, determined by the original mechanical design.

Some companies offer extensions to commercial microscope bodies, e.g. small breadboards that can be installed to replace the standard filter wheels, or vertical breadboards that can be mounted above the sample plane in an inverted microscope [8]. There are also modular microscopy systems available on the market [9] as well as universal opto-mechanical systems, intended, among other applications, for constructing microscope frames [10].

Even a cursory overview of the above mentioned solutions reveals that the designs, commercial and custom-built, tend to be vertical: the optical path in a substantial part of its way travels vertically below and above the horizontally mounted sample. As a result, the opto-mechanics is built in a form of a tower, with optical elements installed in rail or cage systems or, quite often, on vertically mounted breadboards. In optical labs, on the contrary, the majority of experiments are performed on optical tables, where the light beams are sent horizontally, whenever possible on the same height above the table level, throughout the setup. This latter layout is preferred due to laser safety, as the beams are always well below the eye level as well as due to practical reasons. First, the optical elements can be placed on the table and remain there, even before they are permanently attached (clamped, screwed). The beam height above the mounting surface may vary, depending on the experiment design, the optical and opto-mechanical elements used. On the lower extreme are heights on the order of 20-30 mm, with 1/2 inch optics and low-profile mirror and lens mounts. Such a low height guarantees the ultimate mechanical stability and compactness and is often a design of choice in laser design. It is, however, challenging if additional degrees of freedom are to be used, e.g. more than one stage (translation or rotation) mounted in series on top of each other. If more flexibility is required or many large elements are used (e.g. large cameras, gas cells in ovens and/or magnetic shields) the beam height might be up to around 250 mm.

In this paper we present a series of opto-mechanics platforms, developed originally for prototype Raman microscopes: a spontaneous Raman systems in upright configuration with custom-made objective turrets, and two others for inverted Stimulated Raman Scattering (SRS) microscopes, all of them built for imaging biological cells. The general concept in all these designs is to have two horizontal bread boards, connected with a microscope head, where microscope objective(s) and folding mirror(s) are mounted, and which is the only part where light travels vertically. Thus, most of the light beams are aligned horizontally and standard optical and opto-mechanical components can be used, just like on the optical table. The size of the bread boards can be chosen to accommodate the optics on the lower (typically – excitation) and upper (detection) levels. The prototypes were built using many off-the-shelf mechanical components, some of them adapted from commercial microscopes, as well as custom-machined parts of various complexity. The main custom assembly – the microscope head – may be easily adapted for prototype optical microscopes in many different configurations.

**METHODS AND RESULTS**

The departure point for the idea of the opto-mechanics platform was to design it in such a way that as many of the light paths as possible are arranged horizontally, at the height(s) above the horizontal breadboards that would match standard opto-mechanical components, light sources and detectors. The sample plane should be horizontal to allow for water-immersed samples (e.g. cell cultures) and water-dipping objectives. We also wanted to have the objective nosepiece with multiple objective ports, e.g. on a turret. All our prototypes use a galvanometric scanner for point-by-point imaging, but they are designed to be easily adopted for translation stages moving the sample instead.

The first question is: how to move the microscope objective (or the sample) for finding the focus? In commercial microscopes, this is usually done with a rack-and-pinion stage, often with a pair of coarse-fine concentric knobs, conveniently located on the side of the microscope body. Precision rack and pinion stages are rarely used in optical labs, hence most often the vertical (focusing) motion of the objective is realized with standard translation stages installed vertically, with the micrometer screw awkwardly pointing up (or down). Table 1 summarizes possible solutions for the vertical movement of the objective (or the sample) with their respective pros and cons. In our case, designs 1 and 3 use the Z stage with a side micrometer, and design 2 uses a rack-and-pinion stage adapted from a commercial (stereo) microscope with a coarse-fine focusing knobs.

|  | Standard optical translation stage with rack-and-pinion drive | Precision translation stage with rack-and-pinion drive | Precision optical (X) stage with side micrometer | Translation stage with rack-and-pinion drive adapted from a microscope | Precision optical (Z) stage with side micrometer |
|---|---|---|---|---|---|
| **Movement guiding** | Dovetail | Linear ball bearing | Linear ball/roller bearing | Linear ball bearing | Linear ball bearing |
| **Drive** | Rack-and-pinion | Rack-and-pinion | Micrometer screw | Rack-and-pinion | Micrometer screw |
| **Translation precision** | Usually coarse, some coarse/fine | Coarse/fine | Coarse (fine possible with differential micrometers) | Coarse/fine | Coarse (fine possible with differential micrometers) |
| **Range** | Medium | Medium | Large | Medium | Small |
| **Ease of adaptation** | Medium | Medium | Medium | Difficult | Easy |
| **Ergonomics** | Good | Good | Poor | Good | Average/good |
| **Example (\* means that the element was used in one of the prototypes described)** | B05-13, Sugura Seiki *(for the collecting objective in the trans configuration) | FB201 Manual Focusing Block, Prior Scientific | LX10, Thorlabs | T4 Stereo Microscope Coaxial Coarse and Fine Focusing Holder, Wally Sky * | TSD-603, OptoSigma * |

Table 1 | Basic characteristics of the possible translation stages for the vertical (Z, focusing) movement of the objective (or the sample).

The second question is: how to provide a quick, reliable objective exchange? For the first design, of the upright microscope, we developed a series of custom-made "radial" objective turrets, in the rarely used configuration, where the objectives are arranged concentrically in one plane, perpendicular to the turret rotation axis (Figure 2). This is the design of choice for the custom-made objective turret (revolver), as sufficient precision in machining the elements may be achieved, even in a basic workshop. The inverted microscopes, use a commercial objective turret with six ports, available from microscope manufacturers as a separate part (in our case TI2-N-N sextuple nosepiece from Nikon).

The last question is: how to mount the sample to provide a coarse (manual) translation in the XY plane? Here, the upright microscope uses a standard XY stage with micrometers installed on the sides and the sample Z stage mounted on top. Inverted microscopes typically use large, heavy translation stages for the sample positioning. The stage must have a large footprint to provide enough room for the objective exchange from the bottom. This seems to be far from optimum, as a few gram sample (a microscope slide in many cases) is held in place by a component weighting above a kilogram. Our first inverted prototype (design 2) still has a large translation stage, adapted from a commercial inverted microscope, but the last iteration (design 3) has a small XY stage, installed upside-down, with the sample holder extending to the side. This last configuration results in the most compact design of the microscope head, having all the degrees of freedom in the elements' movements, easy access to the sample and the light beams and an ergonomic design. Table 2 summarizes how different solutions have been combined in the three prototypes.

|  | **Design 1** | **Design 2** | **Design 3** |
|---|---|---|---|
| **Configuration** | Upright | Inverted | Inverted |
| **Focusing (moving element)** | Z stage with side micrometer (sample) | Rack-and-pinion (objective) | Z stage with side micrometer (objective) |
| **Objective exchange** | Radial turret (custom-made) | Commercial sextuple nosepiece | Commercial sextuple nosepiece |
| **XY sample translation** | Small XY stage with side micrometers | Large commercial XY stage with rack and pinion drives | Small XY stage with side micrometers, installed upside-down |
| **Beam height** | 35 mm (only the upper level was used) | 72 mm (lower level), 90 mm (upper level) | 62 mm (lower level), 80 mm (upper level) |

Table 2 | Summary of the three microscope platform designs.

## Design 1

The idea of a horizontal microscope was first tested in a prototype shown in Fig. 2 – a simple spontaneous Raman microscope in the epi configuration, with a solid state laser illumination and an external grating spectrometer. The microscope has two base plates: the lower one is made of a 180x32x500 mm long aluminum extrusion (Alutec) and the upper one, mounted on four standard posts, is a 180x400x12 mm solid aluminum plate. The light beams (laser and back-scattered Raman) are guided as low as 35 mm above the upper plate and are directed to/from the microscope objective with a folding mirror (1 inch diameter, silver coated, PF10-03-P01, Thorlabs). The objectives (up to four) are mounted in a custom-made radial turret, presented in different variants in Figure 3.

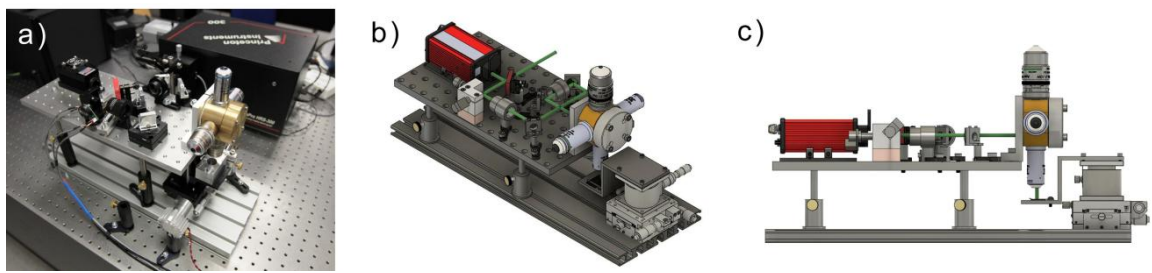

Figure 2 | The upright horizontal microscopes with the custom-made objective turret – a) with the grating spectrometer in the background, b,c) with the green excitation beam visualized. Spontaneous Raman microscope (with the radial turret presented in Fig. 3 a,c) with the Raman excitation laser, the galvo scanner, dichroic mirror, beam guiding mirrors, tube and scan lenses, installed on the upper base plate. Manual XY and Z stages are mounted (stacked) on the lower base plate for focusing and coarse

sample positioning. A grating spectrometer is visible in the background – the Raman signal beam is still delivered to the spectrometer slit via a telescope – but it could be built onto the lower platform as well.

One feature of the radial objective turret with a 45 degree folding mirror is that it can be easily reconfigured to be used for the upright (with the mirror pointing down, as in Fig. 3c) or inverted microscope (with the mirror pointing up, Fig. 3b) by turning the turret body by 180 degrees in its support. In Fig. 3e,f this concept is further developed in an objective head where the optic axis overlaps with the rotation axis of the turret support when the latter is reconfigured between the upright and the inverted position. Thus the configuration can be changed with all the optics remaining in the same position, as the beam height leaving (entering) the turret remains the same. What is more, the turret support with the folding mirror can be rotated to any position if a microscope configuration is needed where the objective axis is not vertical, in particular with the objective axis being oriented horizontally or at any angle with respect to the base plate.

The last configuration – the conical objective turret – comes in two variants: with either one stationary folding mirror or with a number of folding mirrors, each mirror arranged for one microscope objective. The conical objective arrangement allows for the sleek design of the head and the additional advantage is that the unused objectives on the turret do not point upwards, in which case there is often a significant dust accumulation on the front lens surface, even in a clean environment. The designs presented here are intended to demonstrate the radial turret concept and are by no means optimized – detailed solutions will depend on specific applications, space constrains etc.

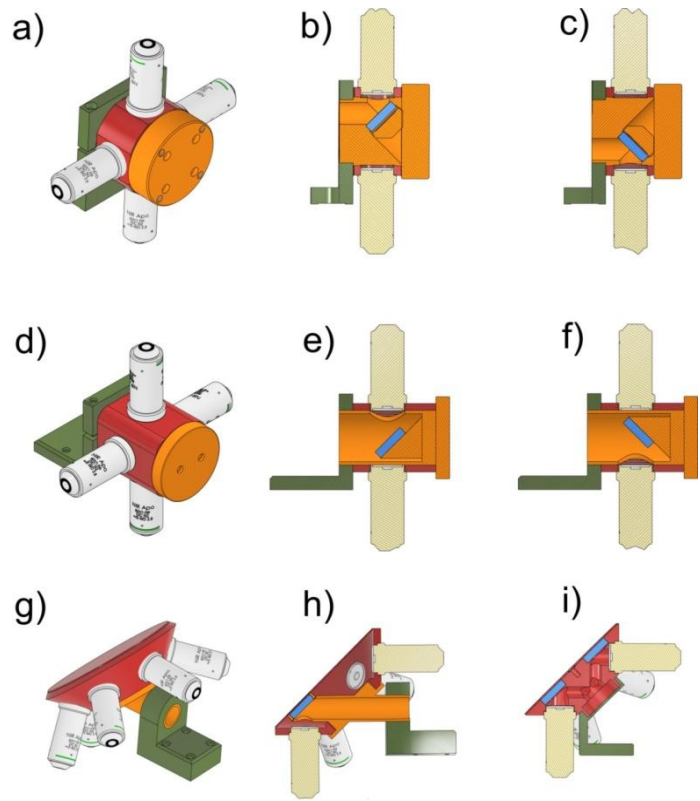

Figure 3 | Different designs of the microscope radial objective turret for compact horizontal microscopes. a) General view of the basic design with four objectives mounted in one plane. b) The

cross section of the design in a) configured for the inverted microscope. c) The cross section of the design in a) configured for the upright microscope. d) General view of the basic design, similar to a), but with the horizontal optic axis overlapping with the axis of rotation of the turret stationary part in the support. e) The cross section of the design in d) configured for the inverted microscope. f) The cross section of the design in d) configured for the upright microscope. g) General view of the conical turret with six objectives. h) The cross section of the design in g) with one mirror used for all the objectives. i) The cross section of the design in g) with each objective using its own mirror. Green: support, yellow: turret stationary part, red: turret revolving part, blue: 45° folding mirror(s).

## Design 2

In this prototype we merged the idea of the light paths being mostly horizontal with the layout of a typical inverted microscope. Two solid supports are mounted on a 300x600x12.7 mm breadboard (MB3060/M, Thorlabs) that is the microscope lower base plate and hold the manual XY stage, adapted from a commercial inverted microscope (Figure 4). The sextuple nosepiece (TI2-N-N, Nikon) is mounted onto a manual coarse/fine focusing block (T4 Stereo Microscope Coaxial Coarse and Fine Focusing Holder, Wally Sky), adapted from a commercial stereo microscope. In this prototype the sample plane is 240 mm above the base (breadboard) plane and the beam height is 72 mm above the lower bread board. The elliptical, 1 inch aperture, silver coated 45 degrees folding mirror, with a special holder (PFE10-P01 and H45E1,Thorlabs) is mounted on a 2-axis kinematic mirror mount, attached horizontally to a standard aluminum post. The infinity space is accessible from approximately 120 mm from the objective parfocal plane.

The upper bread board (180x400x10 mm) has the rack-and-pinion translation stage for the Z movement of the upper objective (25 mm travel range) and a simple mechanism for centering the two objectives: the upper objective is mounted in a conical collar (with the inner thread matching this of the objective, M27x0.75 in our case) that can be translated in the XY plane with two screws (M6/0.5) against the third, spring-loaded screw, the three being mounted every 120 degrees in the XY plane. The upper folding mirror is mounted, without any precision adjustments, to direct the beam 90 mm above the upper bread board, where the detector(s) and an LED for trans illumination are installed as well.

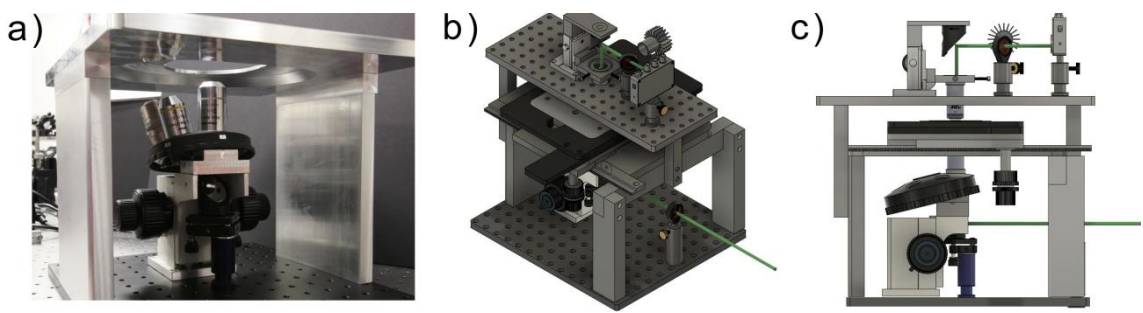

Figure 4 | Design 2 – the first approach to the inverted horizontal microscope. a) In this photo, a large X-Y translation stage is replaced with a solid plate; two spacers compensate for the plate and the stage height mismatch. The focusing (Z) rack and pinion stage is visible with the elliptical folding mirror in a standard kinematic mirror mount (mounted horizontally) on a post. At this stage, the two

supporting plates were located at the sides of the base plate. b,c) The final version with one of the supporting plates moved to the front and the other replaced by two pillars on the sides. It uses a 300x300 mm X-Y translation stage (adapted from Nikon Eclipse). Another three pillars hold the upper breadboard with the upper objective, folding mirror, light detector and LED trans illumination module.

**Design 3**

The last prototype, presented in Fig. 5 is an attempt to make the overall microscope system more compact, to be ultimately integrated in a transportable Stimulated Raman Scattering (SRS) microscope (Fig. 5d). This design combines the idea of folding the beams right before (and after) the objective(s) with one more concept: instead of using a large X-Y translation stage, typically found in inverted microscopes, it relies on a much more compact, lighter X-Y stage, installed upside-down under the upper bread board. A typical inverted microscopes use large XY stages with the central opening that must be wide enough to provide enough space for the objectives (moving on the turret) to approach the sample (usually flush with the stage upper surface) from below. As a result, the stage is typically at least 250x250 mm, with a few cm of travel in each direction, thick (20-30 mm) and heavy. This, in turn, requires a very strong support structure. In our prototype the sample is mounted on a small (80x80 mm, 25 mm travel) XY manual translation stage. The stage is mounted upside-down to the bottom surface of the upper bread board (12 mm thick solid aluminum alloy plate), with the sample support plate (6 mm thick solid aluminum alloy plate) protruding to the side. This way, the small translation stage replaces the large XY translation stage, providing unobstructed access for the objectives from below (the bottom surface of the sample support tray is the lowest element). Focusing is provided with a 10 mm travel Z translation stage (TSD-603, OptoSigma), configured with differential micrometer (MHF2-13F, OptoSigma) and with the same sextuple nosepiece as in design 2. The differential micrometer has 0.5 micron per division fine and 10 micron per division coarse movement, well suited for precision manual focusing. Interestingly, in such configuration, the center of gravity of the objective turret is conveniently located almost exactly above the center of the Z stage platform, thus any torques that could deteriorate the Z stage performance, are avoided. The sample plane is located 165 mm above the base (breadboard) plane and the beam height is 62 mm above the base plane. The infinity space is available from 40 mm from the objective parfocal plane. We also replaced the 2-axis kinematic mirror mount with a monolithic, custom made flexure mount with 45 degrees platform on the moving part to install the lower folding mirror, as its position is only set once. The flexure is used for pitch and the entire mount can be rotated for yaw before being secured to the lower bread board.

The upper bread board is supported with two columns made of aluminum extrusion (90x18.5 mm, Alutec) and has the rack-and-pinion translation stage for the Z movement of the upper objective (20 mm travel range) and the same mechanism for centering the two objectives as in the previous prototype, except it now uses three solid screws, instead of two solid and one spring-loaded ones. Here the upper folding mirror is fixed for 81 mm beam height above the upper base plate. In the later version, the upper folding mirror was mounted on a rack-and-pinion Z stage (20 mm travel range), to allow for adjusting the beam height above the upper base plate between 40-60 mm and the aluminum extrusions have been replaced by solid, 20 mm thick plates to increase the stiffness of the construction (which is always a trade-off with weight). The upper base plate can be larger, if more complex detection setups are to be installed and can be supported with additional (thinner) columns.

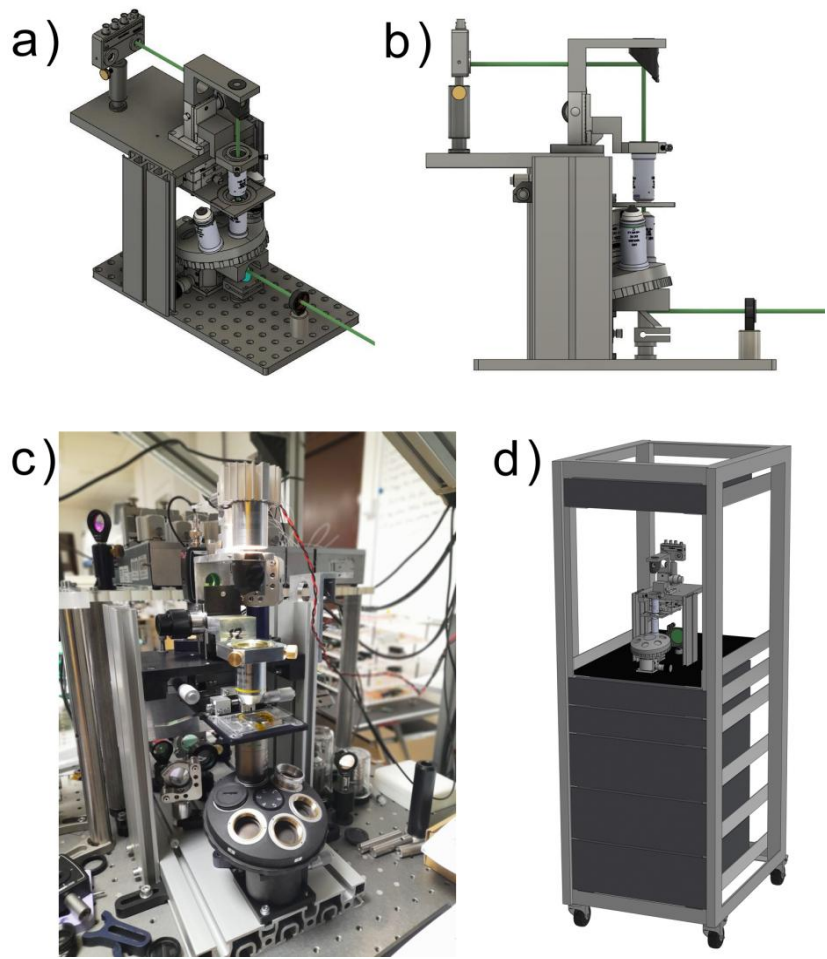

Figure 5 | First version of the Design 3 – a compact inverted horizontal microscope head. It uses a 80x80 mm manual X-Y translation stage (25x25 mm travel range) and the supporting columns are made of aluminum extrusions. The lower folding mirror is installed in a flexure mount with only one degree of freedom (pitch). Design 3 of the platform used for a Stimulated Raman Scattering microscope on the optical table (trans illumination from below with two tunable picosecond fiber lasers (not shown), white LED for wide field trans illumination is also visible on top). Compared to the first variant (a.b), the 180x500 mm base plate is made of an aluminum extrusion and is large enough to accommodate the tube lens, the scan lens, the two-axis galvo scanner and a CMOS camera for wide field sample viewing. The Z stage was also turned by 180 degrees. (d) The mobile, self-contained SRS microscope setup for leukemia cell imaging. Only the microscope head is shown in detail, installed in a standard 19-inch mobile rack with lasers and electronics modules below.

## CONCLUSIONS

Prototype optical microscopes often require specific, non-standard opto-mechanical solutions, beyond those used typically in optics laboratories. We proposed a horizontal microscope platform with a custom-made microscope head that can be used in many configurations and with various components. In a series of prototypes we tested a few approaches to mounting the microscope objectives, the sample and light beam path configurations, using both commercially available, as well as custom-made components. We believe the concepts and designs presented here will be a valuable departure point for the future generation of microscope builders who will adopt them to their specific needs.

## References

[1] Betzig E, Patterson GH, Sougrat R, et al. Imaging intracellular fluorescent proteins at nanometer resolution. Science. 2006;313(5793):1642-1645. doi:10.1126/science.1127344

[2] Tainaka, K., Kubota, S. I., Suyama, T. Q., Susaki, E. A., Perrin, D., Ukai-Tadenuma, M., Ukai, H., & Ueda, H. R. (2014). Whole-Body Imaging with Single-Cell Resolution by Tissue Decolorization. Cell, 159(4), 911–924. https://doi.org/10.1016/j.cell.2014.10.034,

[3] Hell SW, Wichmann J. Breaking the diffraction resolution limit by stimulated emission: stimulated-emission-depletion fluorescence microscopy. Opt Lett. 1994;19(11):780-782. doi:10.1364/ol.19.000780

[4] Huisken J, Swoger J, Del Bene F, Wittbrodt J, Stelzer EH. Optical sectioning deep inside live embryos by selective plane illumination microscopy. Science. 2004;305(5686):1007-1009. doi:10.1126/science.1100035

[5] Dertinger T, Colyer R, Iyer G, Weiss S, Enderlein J. Fast, background-free, 3D super-resolution optical fluctuation imaging (SOFI). Proc Natl Acad Sci U S A. 2009;106(52):22287-22292. doi:10.1073/pnas.0907866106

[6] Rust MJ, Bates M, Zhuang X. Sub-diffraction-limit imaging by stochastic optical reconstruction microscopy (STORM). Nat Methods. 2006;3(10):793-795. doi:10.1038/nmeth929

[7] Owens B. The microscope makers. Nature. 2017;551(7682):659-662. doi:10.1038/d41586-017-07528-7

[8] https://rapp-opto.com/products/microscope-devices-accessories/optical-couplings/ix3-breadboard/#tab-id-2

[9] https://www.thorlabs.com/navigation.cfm?guide_ID=2371

[10] http://www.optoform.com/macroptic-150.html

## Acknowledgements

This work was generously supported within "A platform for fast, label-free imaging, identification and sorting of leukemic cells" project (POIR.04.04.00-00-16ED/18-00) carried out within the TEAM-NET programme of the Foundation for Polish Science, co-financed by the European Union under the European Regional Development Fund. We acknowledge Radosław Bant for preparing the CAD drawings.

**Data availability**

All the data, including detailed drawings of the parts, are available from the corresponding author upon reasonable request.

**Conflict of interest**

P.W. is the named inventor on the patent application P.437476 "Optical microscope" (pending).